\documentclass[a4paper,preprint,showpacs,amssymb,pre,superscriptaddress]{revtex4}
\usepackage{graphicx}

\begin{document}

\title{Bulk Mediated Surface Diffusion: Non Markovian Desorption
with Finite First Moment}

\author{Jorge A. Revelli}\email{jorgerevelli@yahoo.com}
\author{Carlos. E. Budde}\email{budde@famaf.unc.edu.ar}
\author{Domingo Prato}\email{prato@famaf.unc.edu.ar}
\affiliation{Facultad de Matem\'aticas, Astronom\'{\i}a y
F\'{\i}sica, Universidad Nacional de C\'ordoba
\\ 5000 C\'ordoba, Argentina;}
\author{Horacio S. Wio}
\email{wio@ifca.unican.es} \affiliation{Instituto de F\'{\i}sica
de Cantabria, E-39005 Santander, Spain} \affiliation{Grupo de
F\'{\i}sica Estad\'{\i}stica, Centro At\'omico Bariloche and
Instituto Balseiro, \\ 8400 San Carlos de Bariloche, Argentina}

\begin{abstract}
Here we address a fundamental issue in surface physics: the
dynamics of adsorbed molecules. We study this problem when the
particle's desorption is characterized by a non Markovian process,
while the particle's adsorption and its motion in the bulk are
governed by a Markovian dynamics. We study the diffusion of
particles in a semi-infinite cubic lattice, and focus on the
effective diffusion process at the interface $z = 1$. We calculate
analytically the conditional probability to find the particle on
the $z=1$ plane as well as the surface dispersion as functions of
time. The comparison of these results with Monte Carlo simulations
show an excellent agreement.

\end{abstract}

\pacs{46.65.+g, 05.40.FW., 05.10.Ln., 02.50.Eg.}

\maketitle

\normalsize

\section{Introduction}

The mechanism called {\it bulk-mediated surface diffusion}
typically arises at interfaces separating a liquid bulk phase and
a second phase which may be either solid, liquid, or gaseous.
Whenever the adsorbed species is soluble in the liquid bulk,
adsorption-desorption processes occur continuously. These
processes generate surface displacement because desorbed molecules
undergo Fickian diffusion in the liquid's bulk, and are latter
re-adsorbed elsewhere. When this process is repeated many times,
an effective diffusion results for the molecules on the surface.
The importance of bulk-surface exchange in relaxing homogeneous
surface density perturbations is experimentally well established
\cite{c6r1,c6r2,c6r5,c6r9,c6r27}. Adsorption at the solid-liquid
interfaces arises, for instance, in many biological contexts
involving protein deposition \cite{c6r12,c6r13,c6r14}, in
solutions or melts of synthetic macromolecules
\cite{c6r15,c6r16,c6r17,c6r18}, in colloidal dispersions
\cite{c6r19}, and in the manufacture of self-assembly mono- and
multi-layers \cite{c6r20,c6r21,c6r22,c6r23}.

Usually the studies performed in this type of systems are done
within the framework of a Master Equation scheme
\cite{c6r29,c6r30}, where the particle's motion through the bulk
and the adsorption-desorption processes are Markovian. In two
recent papers \cite{c6r31,c6r32} we have shown the most important
features of these phenomena. Non Markovian diffusion processes
have been used in order to modelling a great diversity of
phenomena some of the most relevant are: anomalous charge
transport in amorphous materials \cite{c6r33}, diffusion of
particles with internal states \cite{c6r34}, chromatography
\cite{c6r29}, dielectric relaxation due to defect diffusion
\cite{c6r36}.

We address one of the fundamental issues in surface physics: the
dynamics of adsorbed molecules. We study this problem when the
particle's desorption is characterized by a non Markovian process,
while the particle's adsorption and its motion in the bulk are
governed by a Markovian dynamics. We analyze the diffusion of
particles in a semi-infinite cubic lattice, and focus on the
effective diffusion process at the interface $z = 1$. We calculate
analytically the conditional probability to find the particle on
the $z=1$ plane as well as the surface dispersion as functions of
time, and test those results comparing with Monte Carlo
simulations. When non Markovian processes are present Generalize
Master Equations are involved, these equations being characterized
by a ``memory kernel" that is related univocally with a Continuous
Time Random Walk (CTRW) scheme \cite{c6r37}.

Non Markovian desorption has been introduced in many fields of
natural science describing transport processes in chemical
reactions \cite{c6r38}, re-emission when surfaces contains ``deep
traps" \cite{c6r39}, capture and re-emission from surfaces that
contain sites with many internal states such the ``ladder trapping
model" \cite{c6r40}, proteins with active sites deep inside its
matrix \cite{c6r41} , etc.

The main goal of this work is to get information about the
influence of a non Markovian desorption dynamics on the effective
diffusion process at the interface $z=1$, and in this way to
develop some criteria for looking for non Markovian desorption
effects in experimental situations. For that purpose we calculate
analytically the temporal evolution of the variance ($\langle
r^2(t)\rangle _{plane}$) and the conditional probability of being
on the surface at time $t$ since the particle arrived there at
$t=0$ ($P(x,y,z=1;t|0,0,1,t=0)$), that we indicate by $P(z=1,t)$.

When $\psi(t)$, the waiting time density for desorption defined in
the CTRW scheme, has a {\bf finite first moment} we have been able
to establish the long time asymptotic behavior for $\langle
r^2(t)\rangle _{plane}$ and $P(z=1,t)$ for {\bf any} non-Markovian
desorption. This behavior is the same as the Markovian one and
only depends on the first moment of $\psi(t)$. We also analyzed
the behavior associated to two different families of desorption
waiting time densities. For the first one we have observed two
regions, called {\it strong} and {\it weak adsorption} regime (to
be defined later). In the strong adsorption limit we observe a
transient regime, where the temporal evolution is characterized by
damped oscillations (its frequency not depending on the degree of
departure from Markovian behavior), and a Markovian asymptotic
one. The oscillatory effects disappear in the weak adsorption
region. For the second kind of desorption dynamics, a transient
regime with a non-monotonic behavior for the slope of the variance
emerges, the asymptotic behavior being again Markovian like.

The organization of the paper is as follows. In the next Section
we formally present the model, supported by a Generalized Master
Equation which describe the particle's dynamics through the bulk
and the adsorption-desorption dynamics at the surface. We devote
Section III to present and discuss the time dependence of $\langle
r^{2}(t)\rangle _{plane}$ and $P(z=1,t)$ for two families of
specific memory kernels, and compare the analytical results with
Monte Carlo simulations in the first case, and discuss the general
characteristics in the second. In Section IV we study the
asymptotic behavior of the aforementioned magnitudes, in the
asymptotic long time limit. In the last Section we discuss the
results and present some conclusions.

\section{The adsorption-desorption Model}

Let us start with the problem of a particle making a random walk
in a semi-infinite cubic lattice (with a lattice constant equal to
one). The position of the walker is defined by a vector $\vec{r}$
whose components are denoted by a set of integer numbers $n,m,l$
corresponding to the directions $x$, $y$ and $z$ respectively. The
probability that the walker is at $(n,m,l)$ for time $t$ given it
was at $(0,0,l_0)$ at $t=0$, $P(n,m,l;t | 0,0,l_0,t=0)$ =
$P(n,m,l;t)$, satisfies the following Generalized Master Equation
\begin{eqnarray}
\label{c6mod1}
\dot{P}(n,m,1;t) & = & \gamma P(n,m,2;t)  - \int_0^t dt'\,K(t')
P(n,m,1;t-t'), ~~~~~~~~~~~~~~~~~~~\mbox{for $l=1$} \nonumber \\
\dot{P}(n,m,2;t) & = &  \int_0^t dt'\,K(t') P(n,m,1;t-t') + \gamma
P(n,m,3;t) - \gamma P(n,m,2;t)  \nonumber \\
                 &   & + \alpha
[P(n-1,m,2;t)+P(n+1,m,2;t)-2 P(n,m,2;t)]  \nonumber \\
                 &   & + \beta [P(n,m-1,2;t)+P(n,m+1,2;t)-2 P(n,m,2;t)]
, ~~~~\mbox{for $l=2$} \nonumber \\
\dot{P}(n,m,l;t) & = &  \alpha [P(n-1,m,l;t)+P(n+1,m,l;t)- 2
P(n,m,l;t)]
 \nonumber \\
                 &   & + \beta [P(n,m-1,l;t)+P(n,m+1,l;t)-2 P(n,m,l;t)]
 \nonumber \\
                 &   & + \gamma [P(n,m,l+1;t) +  P(n,m,l-1;t) - 2
                 P(n,m,l;t)], ~~~~\mbox{for $l \geq 3$},
\end{eqnarray}
$\alpha, \beta$ and $\gamma $ are the transition probabilities per
unit time through the bulk in the $x$, $y$ and $z$  directions
respectively and $K(t)$ represents the memory kernel at all sites
$(n,m,l=1)$ and $(n,m,l=2)$ . The form of these equations is
similar to those indicated in \cite{c6r31,c6r32}, with the
desorption parameter $\delta$ replaced by the kernel $K(t)$.

In order to solve the above equations we follow the same procedure
as in \cite{c6r31,c6r32}, taking the Fourier transform with
respect to the $x$ and $y$ variables and the Laplace transform
with respect to the time $t$ in the above equations, we obtain
\begin{eqnarray}
\label{mod2}
s G(k_x,k_y,1;s) - P(k_x,k_y,1,t=0) & = & \gamma G(k_x,k_y,2;s) - K(s)
G(k_x,k_y,1;s),~~~~\mbox{ $l = 1$} \nonumber \\
s G(k_x,k_y,2;s) - P(k_x,k_y,2,t=0) & = & A(k_x,k_y) G(k_x,k_y,2;s)+
K(s) G(k_x,k_y,1;s)+  \nonumber \\
                                    &   & \gamma G(k_x,k_y,3;s)-2 \gamma
G(k_x,k_y,2;s),~~~~~~ \mbox{ $l = 2$} \nonumber \\
s G(k_x,k_y,l;s) - P(k_x,k_y,l,t=0) & = & A(k_x,k_y)
G(k_x,k_y,l;s)+
\nonumber \\
                                    &   & \gamma [G(k_x,k_y,l-1;s)+
\nonumber \\
                                    &   &  G(k_x,k_y,l+1;s)-2
G(k_x,k_y,l;s)].~~~~~ \mbox{ $l \geq 3$}
\end{eqnarray}
We have used the following definitions \cite{c6r31,c6r32}
\begin{eqnarray}
\label{def1}
 G(k_x,k_y,l;s) & = & G(k_x,k_y,l;s |0,0,l_0;t=0) \nonumber \\
& = & \int_{0}^{\infty} e^{-s t}
              \sum_{n,m,-\infty}^{\infty} e^{i(k_x n + k_y m)}
               P(n,m,l;t) dt  \nonumber \\
                             & = & \mbox{} {\it
L}[\sum_{n,m,-\infty}^{\infty} e^{i(k_x n + k_y m)}
                             P(n,m,l;t)],
\end{eqnarray}
where $L$ indicates the Laplace transform of the quantity within
the brackets, and \cite{c6r31,c6r32}
\begin{equation}
\label{def2}
 A(k_x,k_y) = 2 \alpha [\cos(k_x)-1] +
           2 \beta [\cos(k_y)-1].
\end{equation}
Equation (\ref{mod2}) may be expressed in matrix form as
\begin{equation}
\label{mat1}
 [s \tilde{I} - \tilde{H}] \tilde{G} =  \tilde{I},
\end{equation}
where the square matrix $\tilde{G}$ has components
\begin{equation}
\label{mat2} \tilde{G}_{l l_0} = [G[k_x,k_y,l;s | 0,0,l_0;t=0 ]].
\end{equation}
$\tilde{I}$ is the identity matrix and $\tilde{H}$ is a
three-diagonal matrix with the following form
\[\tilde{H}=\left( \begin{array}{cccccc}
 -K(s)  &  \gamma       & 0       & 0      & 0        &
\cdots  \\
  K(s)  &  C            & \gamma  & 0      & 0        &
\cdots  \\
  0     &  \gamma       & C       & \gamma & 0        &
\cdots   \\
  0     &  0            & \gamma  & C      & \gamma   &
\cdots  \\
  \cdot &  \cdot        & \cdot   & \cdot  & \cdot    &
\cdots   \\
                   \end{array}  \right), \]
where $C$ is defined as
\begin{equation}
\label{C} C = -2 \gamma + A(k_x,k_y).
\end{equation}

Above equations for $G(k_x,k_y,l;s)$ are similar to Eq.(2) and
Eq.(5) in \cite{c6r31} with $K(s)$ replacing $\delta$. Therefore,
all results obtained in that paper remains valid for non-Markovian
dynamic when $\delta$ is replaced by $K(s)$ in the Laplace domain.

The expression for the Laplace transform of the variance ($\langle
r^{2}(s)\rangle _{plane}$) became
\begin{equation}
\label{variance4} L[\langle r^2(t)\rangle _{plane}] =\langle
r^2(s)\rangle _{plane}=\frac{N(s)}{D(s)},
\end{equation}
with
\begin{eqnarray}
\label{num} N(s) = & & \left[ 4 K(s) \gamma (\alpha + \beta)
\right]
\\ \nonumber & & \left[ 2 \gamma^3-s^2(\sqrt{s (4 \gamma + s)} -
s) - 3 \gamma^2 (\sqrt{s (4 \gamma + s)}- 3 u)- 2 \gamma s (2
\sqrt{s (4 \gamma + s)} - 3 s) \right],
\end{eqnarray}
\begin{eqnarray}
\label{den} D(s) = & & \sqrt{s (4 \gamma + s)} \\ \nonumber & &
\left[ \gamma s (2 \gamma + s - \sqrt{s (4 \gamma + s)}) + K(s)
(\gamma (\sqrt{s (4 \gamma + s)}- 3 s) + s (\sqrt{s (4 \gamma +
s)}-s))\right] ^{2}.
\end{eqnarray}

The Laplace transform of $P(z=1,s)$, the conditional probability
of being in the surface, is
\begin{eqnarray}
\label{LaplaceP} L[P(z=1,t)] = P(z=1,s)= & & \left[\gamma (2
\gamma + s - \sqrt{s (4 \gamma + s)}\right]  \\ \nonumber & &
\left[ \gamma s (2 \gamma + s - \sqrt{s (4 \gamma + s)}) +
K(s)(\gamma (\sqrt{s (4 \gamma + s)}- 3 s) \right. \\ \nonumber &
& \left. + s (\sqrt{s (4 \gamma + s)}-s))\right]^{-1}.
\end{eqnarray}

Finally we want to point out that the relation between $\psi(t)$,
the waiting time density for the desorption processes defined in
the CTRW scheme, and the memory kernel of Eq. (\ref{c6mod1})
\cite{c6r37} in the Laplace domain is given by
\begin{equation}
\label{KyPsi} K(s)=\frac{s \psi(s)}{(1-\psi(s))}.
\end{equation}

\section{Analytical results and Monte Carlo simulations}

Here we show the results obtained analytically and compare them
with Monte Carlo simulations. In all cases we have fixed the
parameters: $\alpha = \beta = \gamma = 1$.  The simulations
results were obtained averaging over $2 \times 10^{6}$
realizations.

To describe the desorption dynamics from the surface we have
chosen two families of waiting time densities ($\psi(t)$). The
first of them was introduced by Scher and Lax \cite{c6r33} to
describe the frequency dependence of the electric conductivity in
disordered solids when transport is due to impurity hopping. It
has been extensively exploited in modelling non Markovian cases
that emerge when an average over transition rates (disorder) is
taking into account \cite{c6r33} and \cite{c6r37}. The reason of
its wide use are its versatile functional form and its simplicity
which allows to take into account a controllable spread of
transition rates \cite{c6r33}. When only one transition rate is
present the Markovian description is reobtained (the memory kernel
is a Dirac delta function) and when the spread is very wide, steps
occur at fixed regular intervals of time (see below). The adopted
function (see also \cite{new-00}) is
\begin{equation}
\label{pdf}
\psi(t) = \theta a \frac{(\theta a t)^{(a-1)}}
{\Gamma(a)} e^{-\theta a t},
\end{equation}
where $a$ is a positive integer and $\Gamma(a)$ is the Gamma or
Factorial Function. It is worth remarking here two important facts
about this family of functions. First, as can be seen from the Eq.
(\ref{pdf}), there are two parameters which characterized the
function. The parameter $a$, called {\it markovianicity
parameter}, defines the degree of function's departure from the
Markovian behavior ($a=1$ corresponds to the Markovian case; $a
\neq 1$ to the non-Markovian case), while the parameter $\theta$
is the ``average desorption's rate". Second, as shown in
\cite{new-00}, the mean value of these waiting time densities is
\begin{equation}
\label{tmed}
 \langle t\rangle  = \int_{0}^{\infty} t ~~ \psi(t)~~d t = \theta^{-1},
\end{equation}
showing that the ``average desorption's time"  does not depends on
the parameter $a$. For the form of this family of functions, see
Fig. 1 in \cite{new-00}.

\begin{figure}
\centering
\resizebox{.6\columnwidth}{!}{\includegraphics{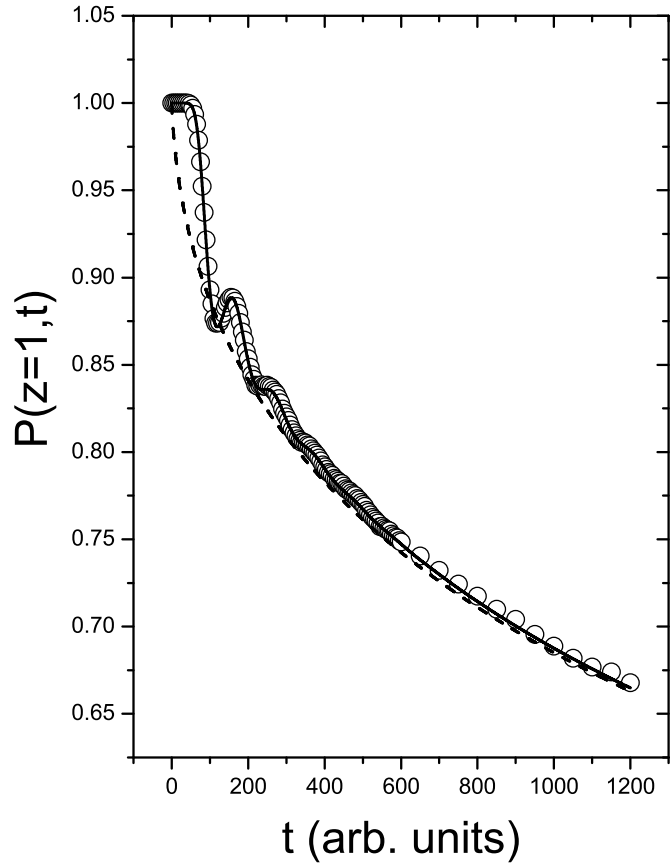}}
\caption{Temporal evolution of the $P(z=1,t)$. We have represented
the case for $\theta = 0.01$. Dot line depicts Markovian
evolution, meanwhile  continuous line and open circles (which
correspond to Monte Carlo simulations) represent the non Markovian
case for $a=20$}.  \label{Fig2}
\end{figure}

\begin{figure}
\centering
\resizebox{.6\columnwidth}{!}{\includegraphics{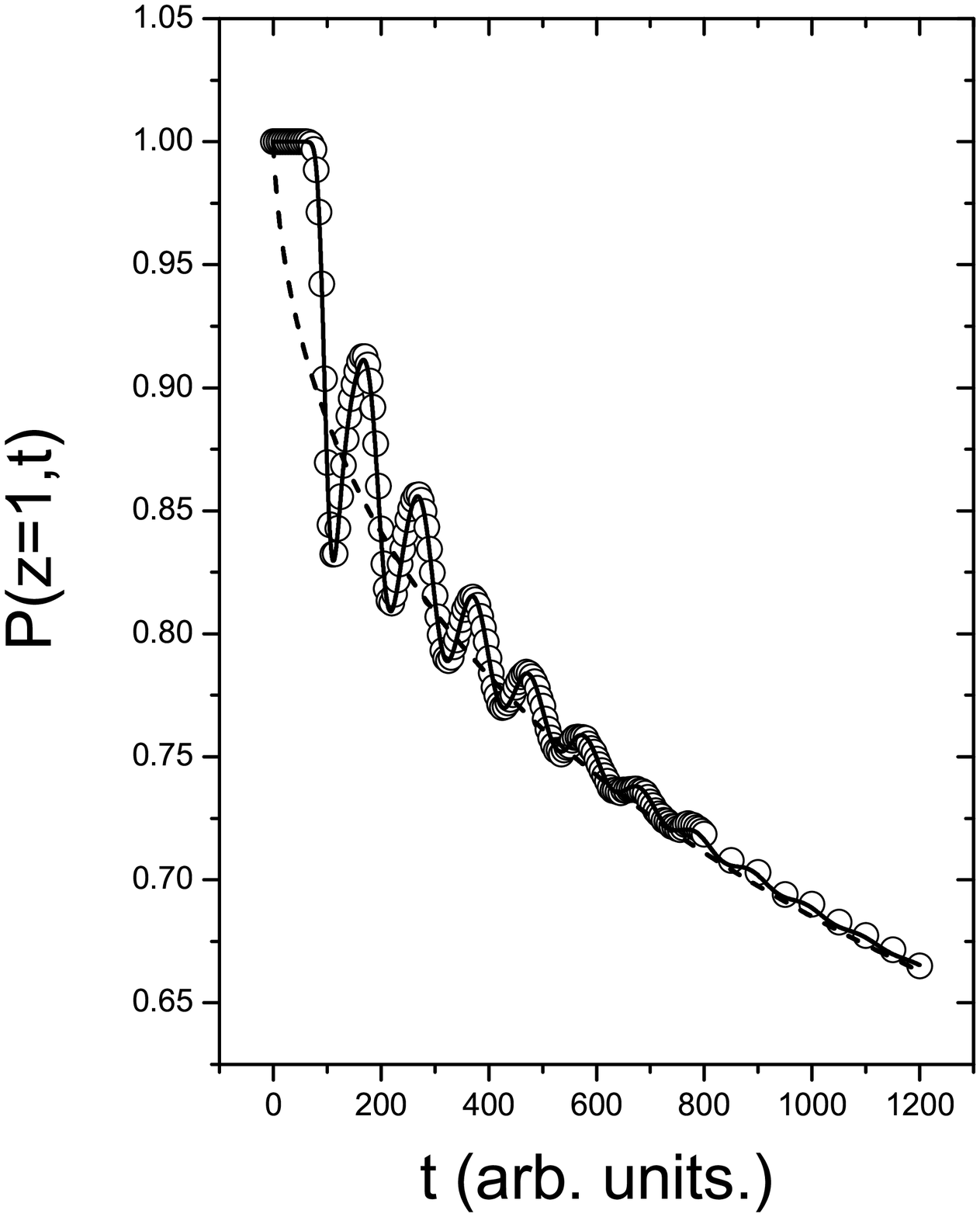}}
\caption{Temporal evolution of the $P(z=1,t)$. We have represented
the case for $\theta = 0.01$. Dot line depicts Markovian
evolution, meanwhile  continuous line and open circles (which
correspond to Monte Carlo simulations) represent the non Markovian
case for $a=75$}.  \label{Fig3}
\end{figure}

In Figs. \ref{Fig2} and  \ref{Fig3} we present the temporal
evolution for the conditional probability ($P(z=1,t)$) for the
Markovian and two non Markovian cases (two different values of
$a$) with the same value of $\theta$. From these figures we can
observe two temporal regions: a transient one which ranges from $t
= 0$ to $t \approx 1000$ and an asymptotic region ($t > 1000$)
where the behavior approaches that of the Markovian case. It is
important to remark the existence of damped oscillations in the
transient region for the non Markovian case. This oscillatory
behavior is due to the waiting time density functions used as will
be explained later.

\begin{figure}
\centering
\resizebox{.6\columnwidth}{!}{\includegraphics{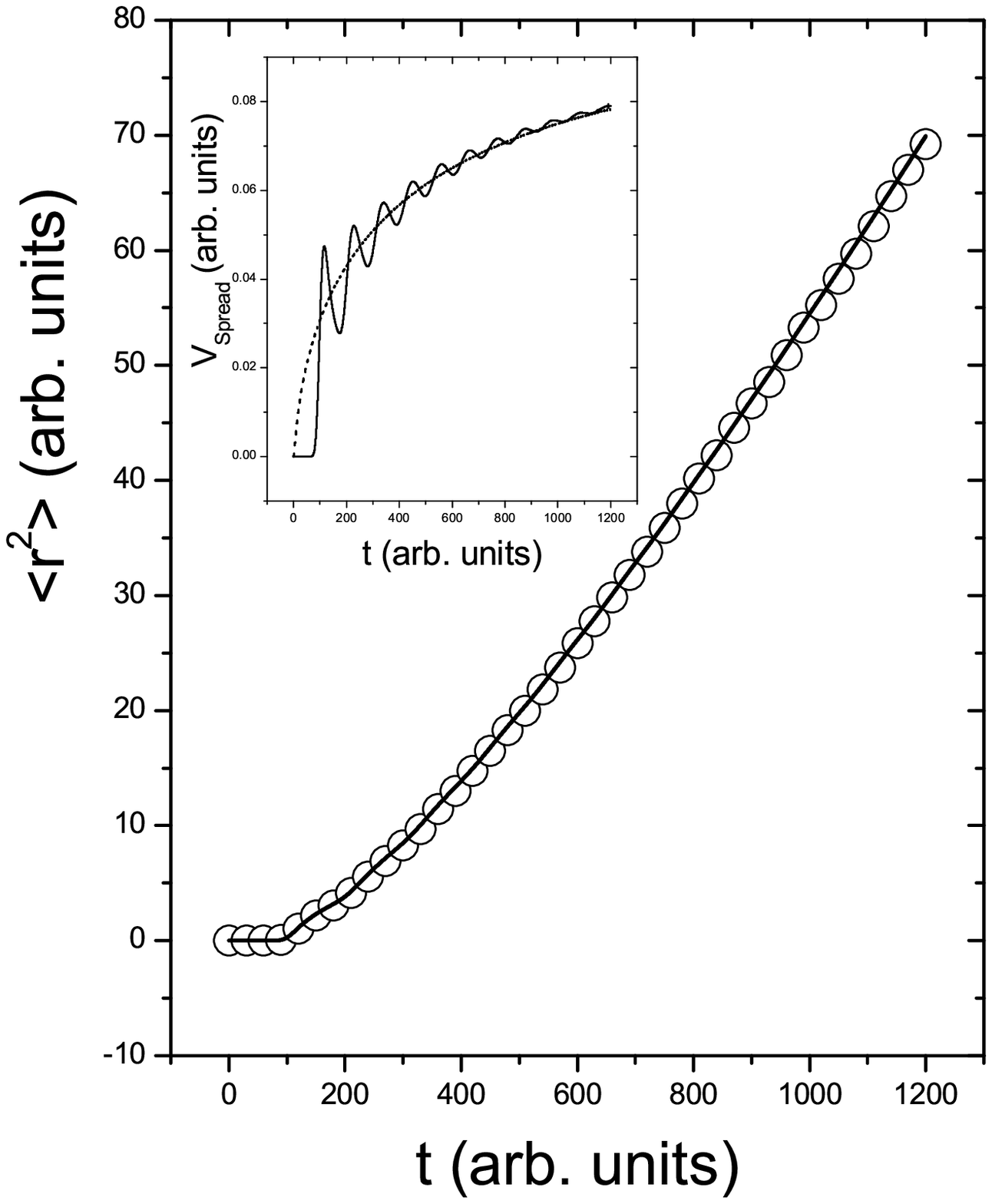}}
\caption{Temporal evolution of the $\langle r^2(t)\rangle
_{plane}$. We have taken $\theta = 0.01$. The solid line and open
circles (which correspond to Monte Carlo simulations) depict the
non Markovian case with $a = 75$. In the insert we can see the
$V_{Spread}$ vs. $t$ for this case (solid line) and the Markovian
evolution (dot line)} \label{Fig4}
\end{figure}

\begin{figure}
\centering
\resizebox{.6\columnwidth}{!}{\includegraphics{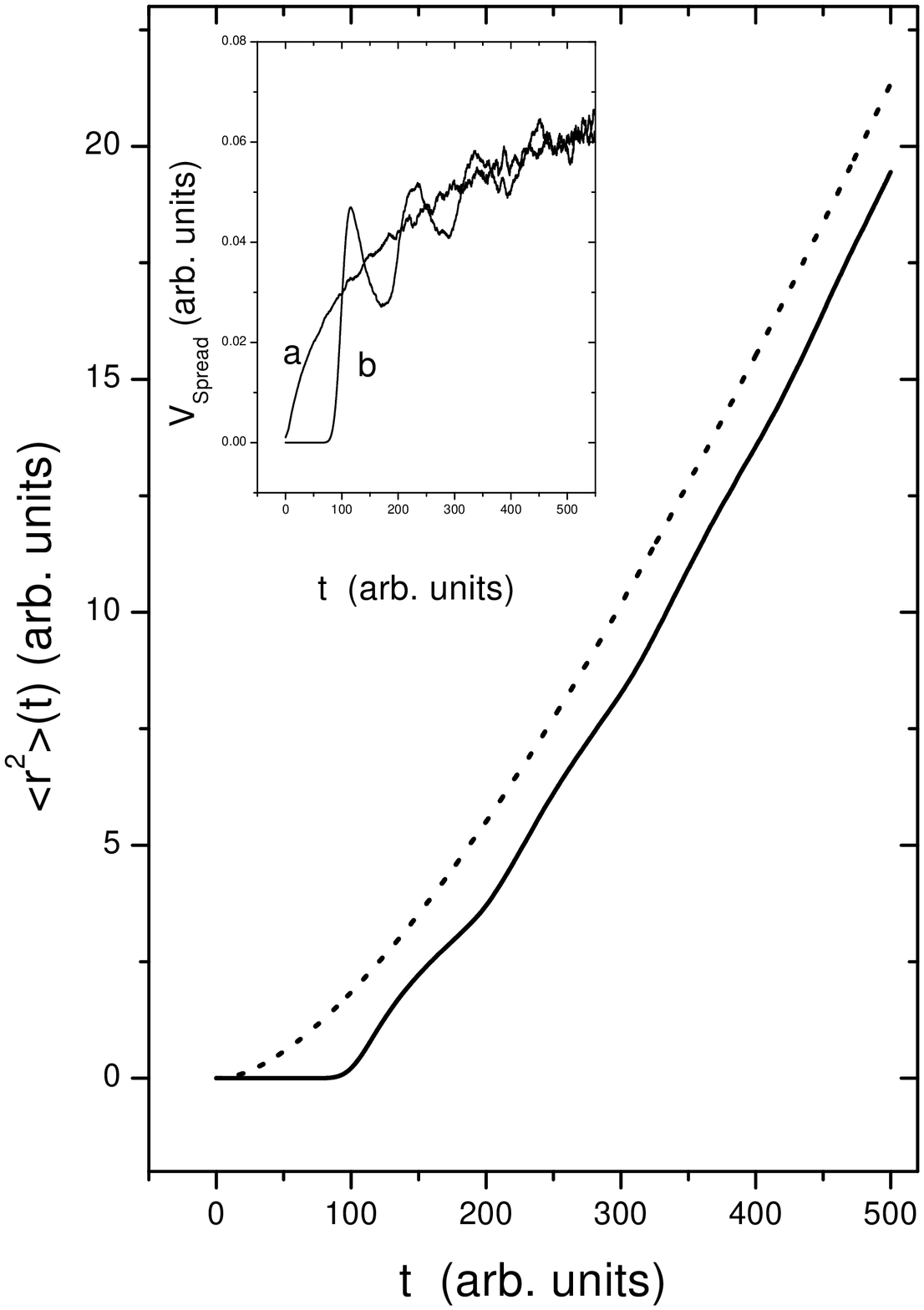}}
\caption{Temporal evolution of the variance $\langle r^2(t)\rangle
_{plane}$. We have represented two cases. Dot line represents the
Markovian case and the continuous line  depict the non Markovian
($a=100$) for $\theta=0.01$. In the insert the curve denoted by
$a$ represents the Markovian case; the $b$ curve represents the
non Markovian one.} \label{Fig5}
\end{figure}

In Figs. \ref{Fig4} and \ref{Fig5}  we depict the results for the
variance $\langle r^2(t)\rangle _{plane}$, for both the Markovian
and the two different non Markovian cases ($a=75$ and $a=100$). We
can observe two important features. The first one is that the
variance in the non Markovian regime shows a delay in the
beginning of the spreading. The second feature is that the system
presents an oscillatory-like behavior, with the oscillations
attenuating as the time grows. This fact is still more apparent in
the insert of the figures where we show the temporal evolution of
the {\it spreading velocity} ($V_{Spread}$). It is worth remarking
here that oscillations only appears in the non Markovian case with
$a \gg 1$ and are due to the particular behavior of the family of
waiting time density defined by Eq. (\ref{pdf}). As it is known,
this function goes to a Dirac delta function as the
markovianicity parameter $a \rightarrow \infty$ ($\psi(t)
\Rightarrow \delta(t-\theta^{-1})$), implying a periodic-like
behavior. This ``periodicity" explains the oscillatory behavior of
$P(z=1,t)$ and $\langle r^2(t)\rangle _{plane}$. Also the figures
show an excellent agreement between the theoretical and Monte
Carlo simulation results.

We remark here that we can distinguish two region defined by the
ratio $\theta / \gamma$. The  {\it strong adsorption region}
characterized by $\frac{\theta}{\gamma} << 1$ where the non
Markovian character shows important differences respect to the
Markovian case in the transient temporal regime and the {\it weak
adsorption region} ($\frac{\theta}{\gamma} > 1$) where these
differences disappear.

\begin{figure}
\centering
\resizebox{.6\columnwidth}{!}{\includegraphics{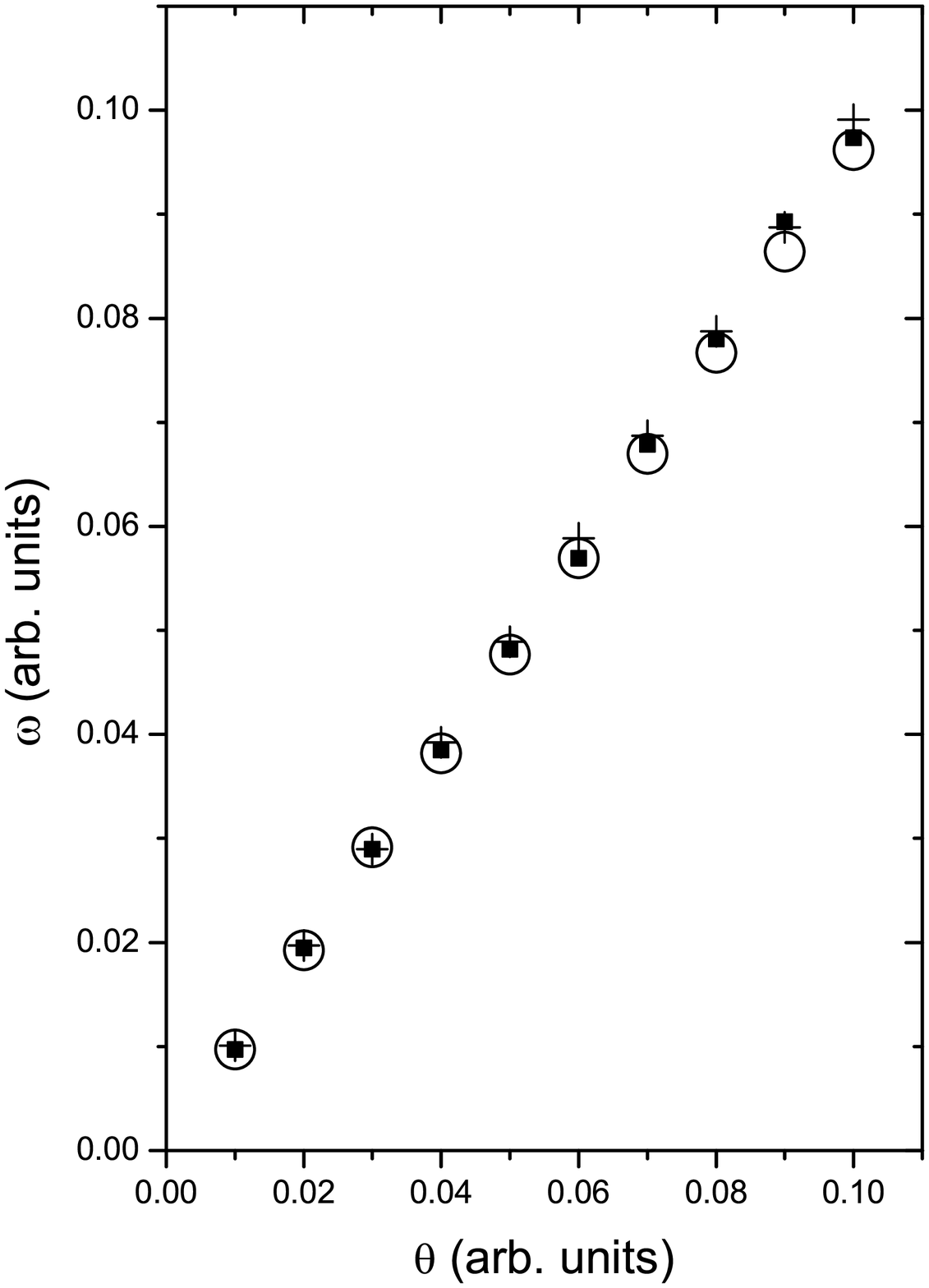}}
\caption{$\omega$ vs. $\theta$. Open circles represents $a = 100$,
squares $a=200$ and cruxes $a=500$.} \label{Fig6}
\end{figure}

Another worth remarking point from these figures is that $\omega$,
the frequency in the oscillation, remains unperturbed due to the
fact that all desorption waiting time densities used have the same
``average desorption's time". This aspect becomes apparent in Fig.
\ref{Fig6}.

\begin{figure}
\centering
\resizebox{.6\columnwidth}{!}{\includegraphics{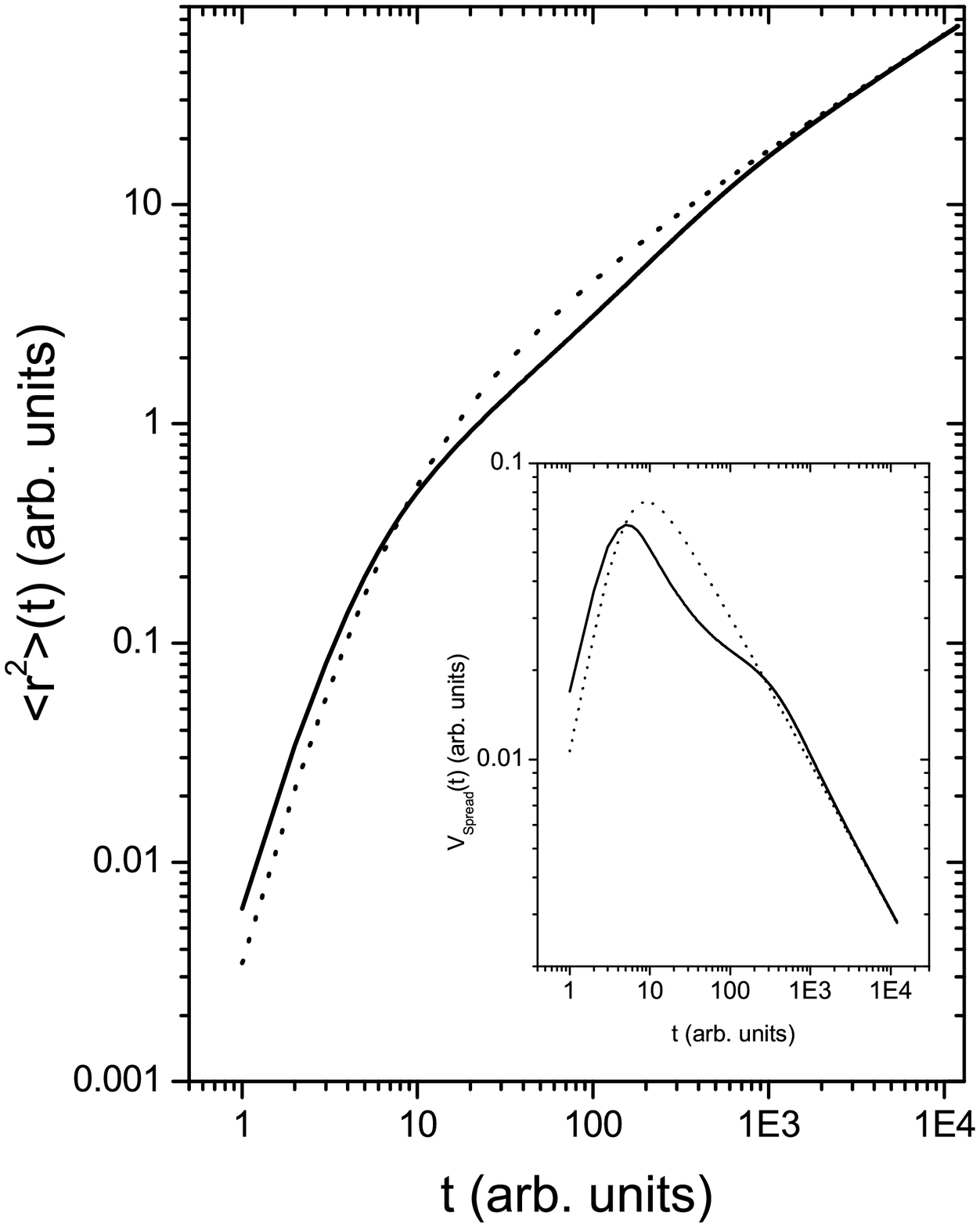}}
\caption{Temporal evolution of the $\langle r^2(t)\rangle
_{plane}$. The solid line depicts the non Markovian case with
$N=2$, $\delta=1$ and $\lambda = 0.01$ and the dot line the
Markovian one. In the insert we can see the $V_{Spread}$ vs. $t$
for this case (solid line) and the Markovian evolution (dot line)}
\label{Fig6b}
\end{figure}

\begin{figure}
\centering
\resizebox{.6\columnwidth}{!}{\includegraphics{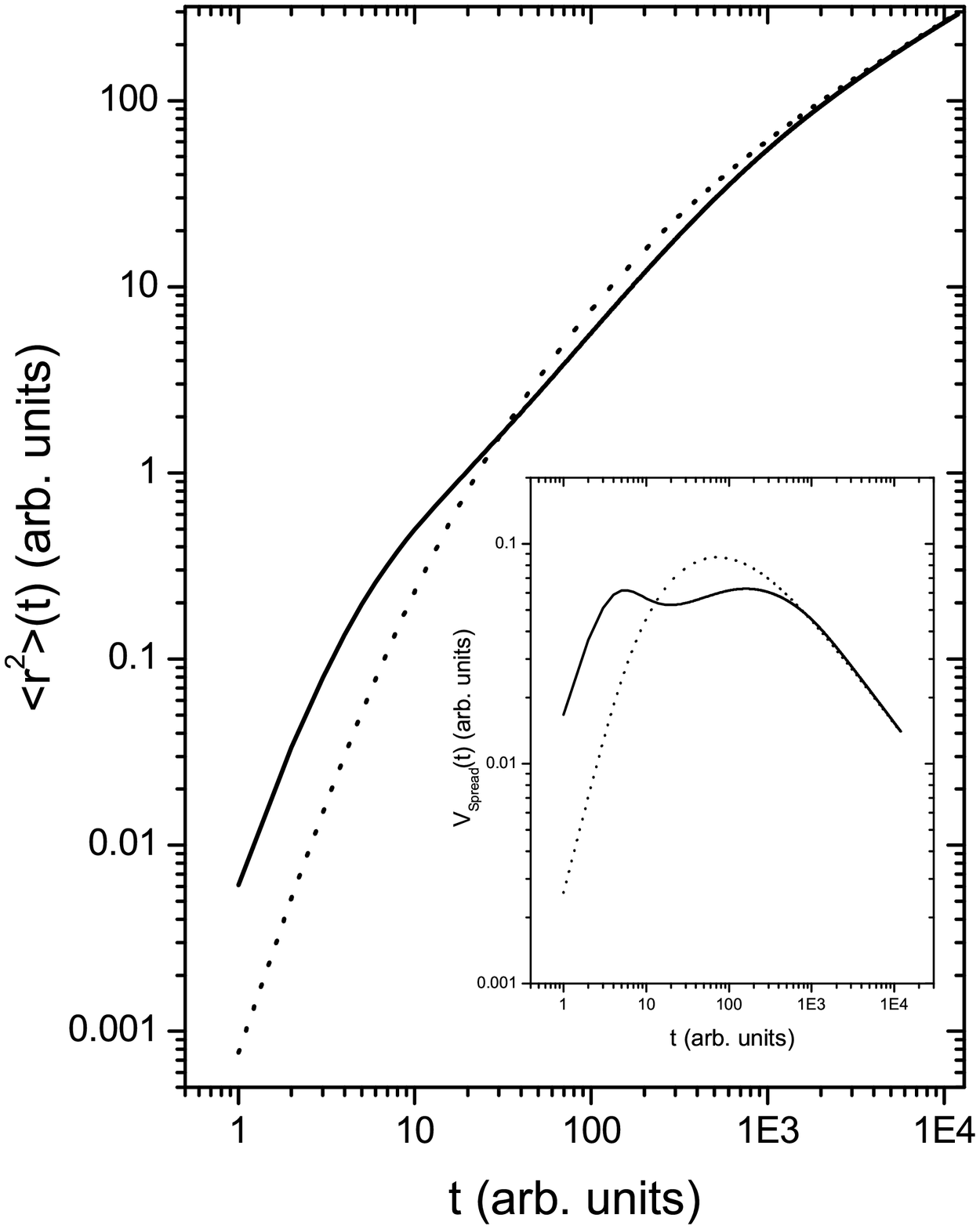}}
\caption{Temporal evolution of the $\langle r^2(t)\rangle
_{plane}$. The same parameters as in Fig. 6 but with $N=10$}
\label{Fig7b}
\end{figure}

The second class of desorption dynamics that we consider in this
work corresponds to the so called ``direct-access multi-trapping
process" \cite{c6r40}. Such an approach was proposed in order to
describe in a phenomenological manner the transport of excited
charge carriers across amorphous material \cite{PyS,Nool}, or the
motion of intersticial in metals \cite{Richter}, etc.  When the
walker arrives at a surface site this model considers transitions
to and from $(N-1)$ internal states, both with probability
$\lambda$ per unit time, before it desorbs to the bulk with
probability $\delta$. The resulting desorption probability waiting
time density in Laplace domain is
\begin{equation}
\label{DiAcc} \psi(s) = \frac{\delta}{\delta+s (1+\frac{\lambda
(N-1)}{s+\lambda})}
\end{equation}
and the ``average desorption's rate" in this case is $\langle
t\rangle  = \frac{N}{\delta}$. For a detailed analysis of this
model and derivation of Eq. (\ref{DiAcc}) see Ref. \cite{c6r40}.
In Figs. \ref{Fig6b} and \ref{Fig7b}  we depict the results for
the variance $\langle r^2(t)\rangle _{plane}$ for two different
non Markovian cases ($N=2$ and $N=10$ respectively). We also show
the Markovian behavior with the same average time for each case
($\langle t\rangle =2$  for $N=2$  and $\langle t\rangle =10$  for
$N=10$). We can observe that the temporal evolution of the
variance in the non Markovian case shows a transient regime
characterized by the slope's non-monotonic behavior, which
increases with $N$. This fact is still more apparent in the insert
of the figures where we show the temporal evolution of the {\it
spreading velocity} ($V_{Spread}$).

\section{Asymptotic behavior}

The asymptotic behavior for large $t$ of $\langle r^2(t)\rangle
_{plane}$ and $P(z=1,t)$ can be obtained  analyzing the limit $s
\ll 1$. We assume that $\psi(t)$, the waiting time density for
desorption, has a {\bf finite first moment} $\langle t\rangle $
which means that $\psi(t)\sim 1-\langle t\rangle s$ when $s\ll 1$
and consequently $K(s)\sim 1/\langle t\rangle $ in this limit.
From  Eqs. (\ref {variance4}), (\ref {num}), (\ref {den}) and
(\ref {LaplaceP}) we obtain
\begin{equation}
\label{asimpru} \langle r^2(t)\rangle _{plane} \sim
\frac{\sqrt{\gamma}(\alpha + \beta)\langle t\rangle }{2
s^{\frac{3}{2}}},
\end{equation}
\begin{equation}
\label{asimpPu} P(z=1,t) \sim \frac{\sqrt{\gamma}\langle t\rangle
}{s^{\frac{1}{2}}}.
\end{equation}

Exploiting Tauberian theorems \cite{c6r37}, the behavior for large
$t$ can be obtained resulting in
\begin{equation} \label{asimprt}
\langle r^2(t)\rangle _{plane} \sim \frac{\sqrt{\gamma}(\alpha +
\beta)\langle t\rangle }{{2\Gamma[3/2]}}
 \,\,\,\,\ t^{\frac{1}{2}},
\end{equation}
\begin{equation}
\label{asimpPt} P(z=1,t) \sim \frac{\sqrt{\gamma}\langle t\rangle
}{\Gamma[1/2]} \,\,\,\,\ t^{-\frac{1}{2}}.
\end{equation}
Equations (\ref{asimprt}) and (\ref{asimpPt}) represent the
behavior of $\langle r^2(t)\rangle _{plane}$ and $P(z=1,t)$ for
{\bf any} non-Markovian desorption for large values of time. This
behavior is the same as the Markovian one and only depends on the
first moment of $\psi(t)$. It is important to remark that the only
assumption was that $\psi(t)$, the waiting time density for
desorption, has a finite first moment $\langle t\rangle $.

\section{Conclusions}

We have studied here the evolution of particles diffusing on a
surface. The diffusion have been performed across the bulk
surrounding the surface, this phenomenon being called {\it bulk
mediated surface diffusion}\cite{c6r31,c6r32}. Usually the
proposed models are based on Markovian desorption processes. The
main feature of this work was to present an analytical model for
non Markovian desorption from the surface. The bulk that surrounds
this surface was considered to be semi-infinite  and that the
particles undergo a Markovian motion on them.

We observed the influence of the non Markovian desorption dynamic
on the effective diffusion process at the interface $z=1$ by
calculating analytically, in the Laplace domain, the temporal
evolutions of the variance ($\langle r^2(t)\rangle _{plane}$) and
the conditional probability of being on the surface at time $t$
since the particle arrived there at $t=0$ ($P(z=1,t)$). When the
waiting time density for desorption has a {\bf finite first
moment} we have established the behavior of the above magnitudes
for {\bf any} non-Markovian desorption for large values of time.
This behavior is the same as the Markovian one and only depends on
the first moment of $\psi(t)$.

We have chosen two families of non-Markovian desorption waiting
time densities and tested the analytical results for $\langle
r^2(t)\rangle _{plane}$ and $P(z=1,t)$ by comparison with Monte
Carlo simulations, obtaining an excellent agreement. For the first
desorption waiting time density we can establish two regions based
on the ratio $\theta / \gamma$, calling them {\it strong
adsorption} (this occurs when $\theta / \gamma << 1$) and {\it
weak adsorption} (when $\theta / \gamma \geq 1$). In the strong
adsorption limit we can observe a transient regime and a
stationary one. The main feature in the transient regime is that
the temporal evolution is characterized by damped oscillations
whose frequencies are in direct relation to  $\theta$, the average
desorption rate. It is worth remarking here that the frequency
does not depend on the markovianicity parameter. The effect of
this parameter appears on the amplitude of the oscillations which
disappear in the weak adsorption region. For the second kind of
desorption dynamics we found a transient regime with an emerging
non-monotonic behavior for the slope's variance. In the asymptotic
regime, the conditional probability  of the non Markovian system
tends to the Markovian one as is expected from the analytical
results obtained in Section IV. In a recent paper \cite{Return} we
have shown that, when we consider {\bf finite} or {\bf infinite
biased} systems, and desorption waiting time densities with finite
first moment, the variance growths linearly with time in the
asymptotic regime.

Finally, it is worth here remarking an important aspect of the
present approach. We have shown through the above results that the
behavior of $\langle r^2(t)\rangle _{plane}$ and  $P(z=1,t)$
results to be strongly dependent on the desorption mechanism. As
the effective dispersion and the percentage of particles that
remain on the plane $z=1$ are measurable magnitudes \cite{c6r30}
they could be used to investigate the characteristic and get
information about the fundamental parameters of the desorption
processes.

\vspace{0.25cm}

{\bf Acknowledgments:} HSW acknowledges the partial support from
ANPCyT, Argentine, and thanks the European Commission for the
award of a {\it Marie Curie Chair}.


\end{document}